\documentclass[preprint,superscriptaddress,notitlepage,endfloat]{revtex4-1}
\usepackage{graphicx,bm,mathtools,color}
\usepackage[pdfstartview=FitH, CJKbookmarks=true, bookmarksnumbered=true, bookmarksopen=true, colorlinks, pdfborder=001, linkcolor=blue, anchorcolor=blue, citecolor=blue]{hyperref}
\usepackage{lineno}
\newcommand{\ket}[1]{\left\vert #1\right\rangle}

\newcommand{\be}{\begin{equation}}
\newcommand{\ee}{\end{equation}}
\newcommand{\ua}{\uparrow}
\newcommand{\da}{\downarrow}
\newcommand{\serge}{CeRh$_{6}$Ge$_{4}$\ }
\begin{document}

\renewcommand{\abstractname}{} 
\title{Strange metal behavior in a pure ferromagnetic Kondo lattice}

\author{Bin Shen}
\thanks{These authors contributed equally to this work}
\affiliation{Center for Correlated Matter and Department of Physics, Zhejiang University, Hangzhou 310058, China}
\author{Yongjun Zhang} 
\thanks{These authors contributed equally to this work}
\affiliation{Center for Correlated Matter and Department of Physics, Zhejiang University, Hangzhou 310058, China}
\author{Yashar Komijani}
\thanks{These authors contributed equally to this work}
\affiliation{Department of Physics and Astronomy, Rutgers University, Piscataway, New Jersey 08854, USA}
\author{Michael Nicklas}
\affiliation{Max Planck Institute for Chemical Physics of Solids, 01187 Dresden, Germany}
\author{Robert Borth}
\affiliation{Max Planck Institute for Chemical Physics of Solids, 01187 Dresden, Germany}
\author{An Wang}
\affiliation{Center for Correlated Matter and Department of Physics, Zhejiang University, Hangzhou 310058, China}
\author{Ye Chen}
\affiliation{Center for Correlated Matter and Department of Physics, Zhejiang University, Hangzhou 310058, China}
\author{Zhiyong Nie}
\affiliation{Center for Correlated Matter and Department of Physics, Zhejiang University, Hangzhou 310058, China}
\author{Rui Li}
\affiliation{Center for Correlated Matter and Department of Physics, Zhejiang University, Hangzhou 310058, China}
\author{Xin Lu}
\affiliation{Center for Correlated Matter and Department of Physics, Zhejiang University, Hangzhou 310058, China}
\author{Hanoh Lee}
\affiliation{Center for Correlated Matter and Department of Physics, Zhejiang University, Hangzhou 310058, China}
\author{Michael Smidman}
\email{msmidman@zju.edu.cn}
\affiliation{Center for Correlated Matter and Department of Physics, Zhejiang University, Hangzhou 310058, China}
\author{Frank Steglich}
\affiliation{Center for Correlated Matter and Department of Physics, Zhejiang University, Hangzhou 310058, China}
\affiliation{Max Planck Institute for Chemical Physics of Solids, 01187 Dresden, Germany}
\author{Piers Coleman}
\email{coleman@physics.rutgers.edu}
\affiliation{Department of Physics and Astronomy, Rutgers University, Piscataway, New Jersey 08854, USA}
\author{Huiqiu Yuan}
\email{hqyuan@zju.edu.cn}
\affiliation{Center for Correlated Matter and Department of Physics, Zhejiang University, Hangzhou 310058, China}
\affiliation{Collaborative Innovation Center of Advanced Microstructures, Nanjing University, Nanjing, 210093, China}

\date{\today}

\begin{abstract}
\textbf{The strange metal phases found to develop in a wide range of
materials near
a quantum critical
point (QCP), 
have posed a long-standing mystery.
The frequent association of strange metals
with unconventional superconductivity and antiferromagnetic QCPs
\cite{QCnphys892,Daou2008,Legros2019,Stewart2001}
has led to a belief that
they are highly entangled quantum
states \cite{Senthil04}.
Ferromagnets, by contrast are regarded as an 
unlikely setting 
for strange metals, for they 
are weakly entangled and their
QCPs are often interrupted by competing phases or first order phase transitions 
\cite{PhysRevLett.82.4707,FMRevModPhys.88.025006,Chubukov2004}. Here,
we provide compelling evidence that the stoichiometric heavy fermion
ferromagnet CeRh$_6$Ge$_4$ \cite{Vosswinkel2012,Matsuoka2015} becomes
a strange metal at 
a pressure-induced QCP: specific heat and resistivity measurements
demonstrate that the FM transition is continuously suppressed to zero
temperature revealing a strange metal phase. 
We argue
that
strong magnetic
anisotropy plays a key role in this process, 
injecting entanglement, in the form of triplet 
resonating valence bonds (tRVBs) into the
ordered ferromagnet. We show that the singular transformation from
tRVBs into Kondo singlets that
occurs at the QCP causes a jump in the Fermi surface volume: a key
driver of strange metallic behavior. Our results open up a new direction for research into FM quantum
criticality, while also establishing an important new setting for 
the strange metal problem.  Most importantly, 
strange metallic behavior at a FM quantum critical point 
suggests that it is  quantum entanglement rather than the destruction of
antiferromagnetism that is the  common driver of 
the many varied examples of
strange metallic behavior.
}
\end{abstract}

\maketitle

Quantum materials augmented by strong electronic correlations are
promising for applications, but the electronic
interactions that empower these materials challenge our
understanding.  One of the most pressing questions in strongly
correlated electronic systems is the origin of  the strange metallic 
behavior which develops at a  quantum critical phase transition between
a delocalized Fermi liquid (FL), and a localized or partially localized electronic phase.  A prime example is the strange metal (SM) phase which develops in the normal state of cuprate superconductors at optimal doping, characterized by
a robust linear resistivity and a logarithmic temperature dependence
of the specific heat coefficient \cite{Daou2008,Legros2019}; similar behavior
is also observed in various quantum critical heavy electron
materials.
The underlying universality of SM behavior that develops in the
vicinity of  QCPs is currently a subject 
of intense theoretical
interest. 
One of the valuable ways
of identifying the key ingredients of SM behavior is through experiments 
that explore new classes of quantum materials. 

Kondo lattice systems with periodically arranged atoms hosting localized $f$-electrons show a rich variety of properties,
due to competition between magnetic interactions among local 
moments, and their ``Kondo'' screening by conduction
electrons \cite{QCnphys892}.   The small energy scales of
these interactions leads to highly tunable ground states, which is ideal for  studying
SM behavior. In a number of
systems, the tuning of the 
aforementioned competition leads to a continuous suppression of 
antiferromagnetic (AFM) order at a quantum critical point (QCP)\cite{Stewart2001}. 
The outcome when a ferromagnetic (FM)
transition is suppressed by a non-thermal tuning parameter is
generally different \cite{FMRevModPhys.88.025006}. FM
QCPs are often avoided by the occurrence of a first order
transition \cite{PhysRevB.55.8330}, the intersection of AFM phases
\cite{Sullow1999,PhysRevLett.101.026401}, or a Kondo cluster glass phase
\cite{SGPhysRevLett.102.206404}. This raises the question of whether
AFM correlations are crucial for realizing SM behavior.

Early theoretical studies of 
itinerant ferromagnets\, \cite{PhysRevLett.82.4707,Chubukov2004} in
the framework of Hertz-Millis-Moriya (HMM) theory\ \cite{Sachdev2011}
predicted that quantum phase transitions in these materials
are inevitably driven first order by interactions 
between  the critically scattered electron
fields, thereby interrupting 
the development of  quantum criticality. 
However, the recent discovery of a
FM QCP in the heavy fermion system
YbNi$_4$P$_2$ tuned by chemical pressure \cite{Steppke2013}, 
 raised the fascinating
possibility that the FM QCP in 
these systems 
is governed by a different universality class,  involving a
break-down of  Kondo screening \cite{YRSnature,407nature,Yamamoto2010}. 
The negative
pressure required to reach the FM QCP of YbNi$_4$P$_2$
necessarily involves chemical doping of the stoichiometric compound,  which  introduces
disorder, complicating the theoretical interpretation. Disorder suppresses first order transitions  \cite{PhysRevLett.82.4707}, as in the case of ZrZn$_2$, where early experiments suggested the presence of a
FM QCP  \cite{Smith1971}, but improved sample quality led to a first order transition \cite{PhysRevLett.93.256404}.  Thus while the experimental
data on YbNi$_4$P$_2$  suggests the existence of FM QCPs, 
a definitive proof of such behavior in  a quantum ferromagnet requires utilizing hydrostatic, rather than chemical pressure. 
Ce-based heavy fermion ferromagnets, where pressure can cleanly tune
the system to a QCP,  are ideally suited for such
studies.

CeRh$_6$Ge$_4$ is a heavy fermion ferromagnet with  a Curie temperature    $T_{\rm C}=2.5$~K \cite{Matsuoka2015}. The crystal structure (Fig.~1a) consists of  triangular lattices of Ce  stacked along the $c$-axis \cite{Vosswinkel2012}. The Ce-Ce separation is much smaller  along the $c$-axis (3.86~\AA) than in the triangular planes (7.15~\AA), suggesting a  quasi-one-dimensional nature to the magnetism. Under hydrostatic pressure, we find that the FM transition of CeRh$_6$Ge$_4$ is smoothly suppressed to zero temperature,  reaching a QCP at $p_c=0.8$~GPa. 

The temperature dependence of the resistivity $\rho(T)$ and specific heat (as $C(T)/T$) of single crystalline CeRh$_6$Ge$_4$ both show   transition anomalies at around $T_{\rm C}\approx2.5$~K (Figs.~1b and 1c). When magnetic fields are applied within the $ab$-plane, the transition becomes a broadened crossover, consistent with FM ordering. The low temperature magnetization (as $M/H$) is shown in Fig~1d. Measurements up to 300~K demonstrate that the magnetic easy direction lies within the $ab$-plane (Fig.~S1). On cooling, just above $T_{\rm C}$, the in-plane $M/H$ undergoes a marked enhancement, typical of FM order.  For fields along the $c$-axis, $M/H$ abruptly increases at the transition. Magnetization loops below $T_{\rm C}$ for in-plane fields show  hysteresis characteristic of FM materials (Fig.~1e). At higher fields, there is no hysteresis, and a much less rapid increase of the magnetization with field above 0.28$\mu_{\rm B}$/Ce, which likely corresponds to the ordered moment (Fig.~S1). 

The zero-field resistivity and specific heat coefficient at various pressures are displayed in Figs.~2a and 2b, respectively (see also Figs.~S3 and S4). The evolution of the properties with pressure and  the resulting $T-p$ phase diagram are presented in Figs.~3a and 3b. At $T_{\rm C}$ the resistivity crosses over from a $T$-linear  behavior at high temperature to a $T^2$ behavior at low temperatures (Fig.~S3), where $C(T)/T$ becomes temperature independent. The FM transition, which is suppressed almost linearly by pressure, cannot be detected anymore beyond $p_c=0.8$~GPa.  In the  paramagnetic (PM) phase above $p_c$, the aforementioned low-$T$ FL properties are again observed  (Figs.~S3 and S4). The temperature at which this FL behavior onsets ($T_{\rm FL}$) increases almost linearly with pressure (Fig.~3b). Both the value of the low-temperature plateau in $C(T)/T$ and the $A$-coefficient in $\rho(T)=\rho_0+AT^2$ show an incipient divergence when approaching $p_c$ from the FM or PM side (Fig.~3a). On both Fermi-liquid sides of the phase diagram, the  Kadowaki-Woods ratio $A/\gamma^2$  is $1.49\times10^{-6}$ (ambient pressure) and $1.33\times10^{-6}\mu\Omega$~cm~mol$^2$~K$^2$~mJ$^{-2}$ (1.12~GPa), which are close to the value for a $4f$-electron ground state degeneracy $N=4$.

At $p_c=0.8$~GPa, the resistivity is strictly linear in temperature over two decades down to at least 40~mK, while $C(T)/T \propto {\rm log}(T^*/T)$ with { $T^*=2.3$~K ($T^*$ is a characteristic temperature of  the spin fluctuation energies \cite{Stewart2001})}, over more than a decade (Fig.~2c). At  60~mK, $C(T)/T$ reaches a very large value of 1.1~J~mol$^{-1}$K$^{-2}$. Between the FM and PM phases, there is  a fan-shaped SM region with properties similar to canonical AFM quantum critical systems such as CeCu$_{6-x}$Au$_x$ \cite{Lohneysen1994} and YbRh$_2$Si$_2$ \cite{Trovarelli2000}.  The pressure dependences of  the $A$-coefficient  and the Sommerfeld coefficent $\gamma$ (Fig.~3a) follow the residual resistivity $\rho_0$, which also develops a maximum at $p_{\rm c}$, reflecting the presence of strong quantum critical fluctuations (Fig.~S3).
 
At first sight, the strange metal properties of
\serge might be attributed to itinerant quantum criticality, for
aside from the absence of a first order phase transition, HMM
theory predicts a logarithmic Sommerfeld coefficient 
and a $T$-linear electron scattering rate, naively equivalent to
a $T$-linear resistivity\ \cite{Stewart2001}. However, 
the scattering off long-wavelenth FM fluctuations does
not relax electron currents,  
and once this effect is included, a $\rho\sim T^{5/3}$ dependence of the
resistivity is expected \cite{Stewart2001}.
A $T$-linear
resistivity suggests large angle scattering, a feature
 typical of {\it local} fluctuations. 
Moreover, the strength
of the logarithmic divergence in the specific heat anomaly,
determined by the fit
$C/T\sim \frac{S_{0}}{T^{*}} \log(T^{*}/T)$, shows that the entropy 
$S_{0}\sim
\frac{1}{10} R \log(2)$ is a large fraction 
of the local moment entropy, once again suggesting an underlying local
mechanism to the quantum criticality. Together with the absence
of a first order phase transition, these features 
provide strong evidence in favor of a local
quantum critical point.

In AFM heavy electron
metals, the development of a $T$-linear resistivity 
coincides with an abrupt jump in the Fermi surface volume, 
accompanied by singular charge
fluctuations \cite{Paschen04,Shishido05,Komijani19}. It has been
argued that such a jump in the Fermi surface is caused by an abrupt
transformation in the pattern of spin entanglement
\cite{Senthil04}, as the Kondo singlets transform into resonating
valence bonds (RVBs) in the spin fluid.
This leaves us with a puzzle, for the spins in a simple ferromagnet
are not entangled,
which would imply a \emph{continuous} evolution 
of the Fermi surface
\cite{Komijani18}. The unusual aspect  of CeRh$_6$Ge$_4$ is the
development of a SM phase at a FM QCP, which shows similar behavior
to the non-stoichiometric material
YbNi$_4$P$_{2-x}$As$_{x}$\,\cite{Steppke2013}. 

Apart from the
quasi-1D nature, a common feature of these two materials is an
easy-plane anisotropy. In such systems, the magnetic order parameter
is no longer conserved and will develop marked zero-point
fluctuations, likely responsible for the severely reduced magnetic
moment.  This can be seen clearly in a
two-site example where the magnetization is along the
$x$-direction. The ordered phase is a product state which can be expanded in
terms of
triplets, 
\be \Big(\frac{\ket{\ua_i}+\ket{\da_i}}{\sqrt
2}\Big)\Big(\frac{\ket{\ua_j}+\ket{\da_j}}{\sqrt
2}\Big)=\frac{\ket{\ua_i\ua_j}+\ket{\da_i\da_j}}{2}+\frac{1}{\sqrt{2}}\Big(\frac{\ket{\ua_i\da_j}+\ket{\da_i\ua_j}}{\sqrt
2}\Big).  \ee 
An easy-plane  anisotropy 
projects out the equal-spin pairs on the right-hand-side, creating a
triplet valence bond. 
In a lattice, the same effect 
creates a quantum superposition of triplet pairs, forming
a triplet-RVB (tRVB) state, written schematically 
$P_G\ket{\rm FM}=\ket{\rm tRVB}$.  Hence, 
an easy-plane anisotropy in FM systems plays
the same role as magnetic frustration in AFM systems, injecting
a macroscopic entanglement into the ground state.  
This leads
us to hypothesize that the SM behavior at the FM QCP has its origins
in the magnetic anisotropy.

To test these ideas, we have studied 
a simplified Kondo lattice model with nearest neighbor FM couplings
with
an 
easy-plane anisotropy of the form
$-J^{ij}_{xy}(S^x_iS^x_j+S^y_iS^y_j)-J^{ij}_z S^z_iS^z_j$ on a tetragonal lattice,
consisting of spin chains along the $c$-direction with weak
inter-chain couplings (see Supplementary Material). 
When the chains are weakly coupled, our
simulations indicate the development of a second order phase
transition, while at higher couplings a first order phase transition
develops. This feature is 
in agreement with the current observations of FM QCPs developing in
quasi-1D systems. We assume $J_{xy}>J_z$ which has a dual effect:
it
converts the model into an easy plane $x-y$ ferromagnet, 
and generates triplet resonating valence bonds (tRVB).
Also, the  anisotropy changes the magnetic 
dispersion at low momenta from  quadratic to linear (see Supplementary Material).
By switching on the Kondo screening
\cite{Komijani18,Komijani19,Wang19} we can then tune the model to the 
QCP. 
 
Our calculations take advantage of a Schwinger boson 
representation of the magnetic moments which allows us to treat 
the magnetic and Kondo-screened parts of the  phase
diagram, and the QCP that links them together  (Fig.~3c). 
The key feature of this approach, is that the Kondo effect causes the
spins to fractionalize into electrons, expanding the Fermi surface
while leaving behind  positively
charged, spinless,  Kondo singlets.
In the ordered phase, the majority
of the moments are aligned, 
while some  form tRVB pairs with their
neighbors. 
In an isotropic ferromagnet, the continuous growth of magnetization away from the QCP, indicates a continuous change in the fraction of Kondo screened moments,
or a continuous evolution of the Fermi surface. 
However,  when the moments entangled within tRVB states are abruptly released into the Fermi sea, we find (Supplementary text)  that there is a jump in the Fermi surface volume. The resulting QCP is a plasma, in which the Kondo singlets, the
electrons and the RVB bonds are in a state of critical dynamical
equilibrium, giving rise to singular spin and charge fluctuations as well as a logarithmic in temperature specific heat coefficient (Supplementary text), in agreement with our experimental results.

Our findings of a pressure-induced QCP in CeRh$_6$Ge$_4$ demostrate that
a FM system can develop a continuous quantum
phase transition in the absence of disorder, a result that at present,
can only be
understood in the framework of local quantum criticality, where Kondo screening is suppressed to
zero at the QCP. The observation of SM behavior at finite temperatures above the
QCP with  a $T$-linear resistivity and a specific heat coefficient
that is logarithmically divergent in $T$, 
now expands the scope of this
phenomenon to encompass ferromagnets. 
Central to the SM behavior in a ferromagnet 
is a small abrupt jump in the Fermi surface volume. {An
experimental observation of such a jump would be an unambiguous 
test of Kondo breakdown, as there is no unit-cell doubling at a FM
phase transition. }

Finally, spin-triplet superconducting pairing states have been
proposed in FM heavy-fermion systems,  such as UGe$_2$ \cite{Saxena00}
and URhGe \cite{Levy05}. While there is no sign of superconductivity
in \serge down to 40mK, it is very likely that at sufficiently low
temperatures, the tRVB states that are already present in the critical
regime will migrate into the conduction band as a triplet superconducting
condensate.

\clearpage

\noindent \textbf{Methods} \\
\noindent\textbf{Crystal growth and characterization.} Needle-like shaped single crystals of CeRh$_6$Ge$_4$ were grown using a Bi flux \cite{Vosswinkel2012}. The elements were combined in a molar ratio of Ce:Rh:Ge:Bi of 1:6:4:150, and sealed in an evacuated quartz tube. The tube was heated and held at 1100$^{\circ}$C for 10 hours, before being cooled at 3$^{\circ}$C/hour to 500$^{\circ}$C. The tube was then removed, and centrifuged to remove the excess Bi. The orientation of the crystals was determined using  single crystal x-ray diffraction, and the chemical composition was confirmed using energy dispersive x-ray spectroscopy. The samples measured under pressure had typical values of $\rho_0\approx1.6$~$\mu\Omega$~cm and $RRR=\rho(300~\rm{K})/\rho(0.3~\rm{K})\approx45$ (Fig.~S2). 

\noindent \textbf{Physical property measurements.} Magnetization measurements were  performed using a Quantum Design Magnetic Property Measurement System (MPMS).  The heat capacity at ambient pressure was measured  down to 0.4~K, in applied magnetic fields up to 14~T, using a Quantum Design Physical Property Measurement (PPMS) system with a $^3$He insert, utilizing the  standard relaxation method. Specific heat experiments under pressure were carried out using a CuBe piston-cylinder-type pressure cell \cite{Nicklas2015}. The sample and a piece of lead as pressure gauge were put in a teflon capsule together with Flouinert serving as liquid pressure transmitting medium. The capsule was then mounted inside the pressure cell. The heat capacity of the whole assembly was determined by a compensated heat-pulse method in a dilution refrigerator (Oxford Instruments) down to temperatures of 60~mK. To obtain the heat capacity of the sample the addenda has been recorded in a separate measurement run and subtracted for each pressure from the data obtained of the whole setup including the sample. The pressure inside the cell was determined by the pressure-induced shift of the superconducting transition temperature of the piece of lead measured in a Quantum Design MPMS. The  magnetic field was removed in an oscillating fashion to reduce the remanent field ($<3$~Oe) of the superconducting magnet. The remaining effect on the superconducting transition temperature was compensated for by determining the shift of the superconducting transition of the lead inside the pressure cell with respect to a reference piece fixed to the outside. Electrical transport and ac calorimetry measurements under pressure were carried out in a piston-cylinder clamp-type cell with  Daphne oil 7373 as a pressure transmitting medium. The pressure was also determined from the superconducting transition of Pb.  The resistivity was measured using the four contact configuration between 0.05~K and 300~K. The measurements between 1.9~K and 0.4~K were performed {in a $^3$He refrigerator}.

\noindent \textbf{Data availability} All the data supporting the findings are available from the corresponding author upon reasonable request.
\\
\\
\textbf{Acknowledgments} We would like to thank Cornelius Krellner and Manuel Brando for fruitful discussions,  Guanghan Cao and Zhicheng Wang for assisting with $^3$He-SQUID measurements, and Xiaoyan Xiao for  assistance with single crystal x-ray diffraction. This work was supported by the National Key R\&D Program of China
(No.~2017YFA0303100, No.~2016YFA0300202), the National Natural Science
Foundation of China (No.~U1632275), the Science
Challenge Project of China (No.~TZ2016004) and the National Science
Foundation of the United States of America,
grant DMR-1830707.\\
\\
\textbf{Additional information} Correspondence and requests for materials should be addressed to H. Q. Yuan (hqyuan@zju.edu.cn), P. Coleman (coleman@physics.rutgers.edu) or M. Smidman (msmidman@zju.edu.cn).
\\
\\
\textbf{Author contributions} The project was concieved by  H. Y.. The crystals were grown by Y. Z. and H. L.,  and measurements were performed by B. S., Y. Z., M. N., R. B., A. W., Y. C., Z. N., R. L., and X. L.. The experimental data were analyzed by B. S., Y. Z., M. N., H. L., M. S., F. S. and H. Y.. Theoretical calculations were performed by Y. K. and P. C.. The manuscript were written by Y. K., M. S., F. S., P. C., and H. Y. All authors participated in discussions.
\\
\\
\textbf{Competing financial interests} The authors declare no competing financial interests.

\clearpage

\begin{figure}[h]
\begin{center}
 \includegraphics[width=0.7\columnwidth]{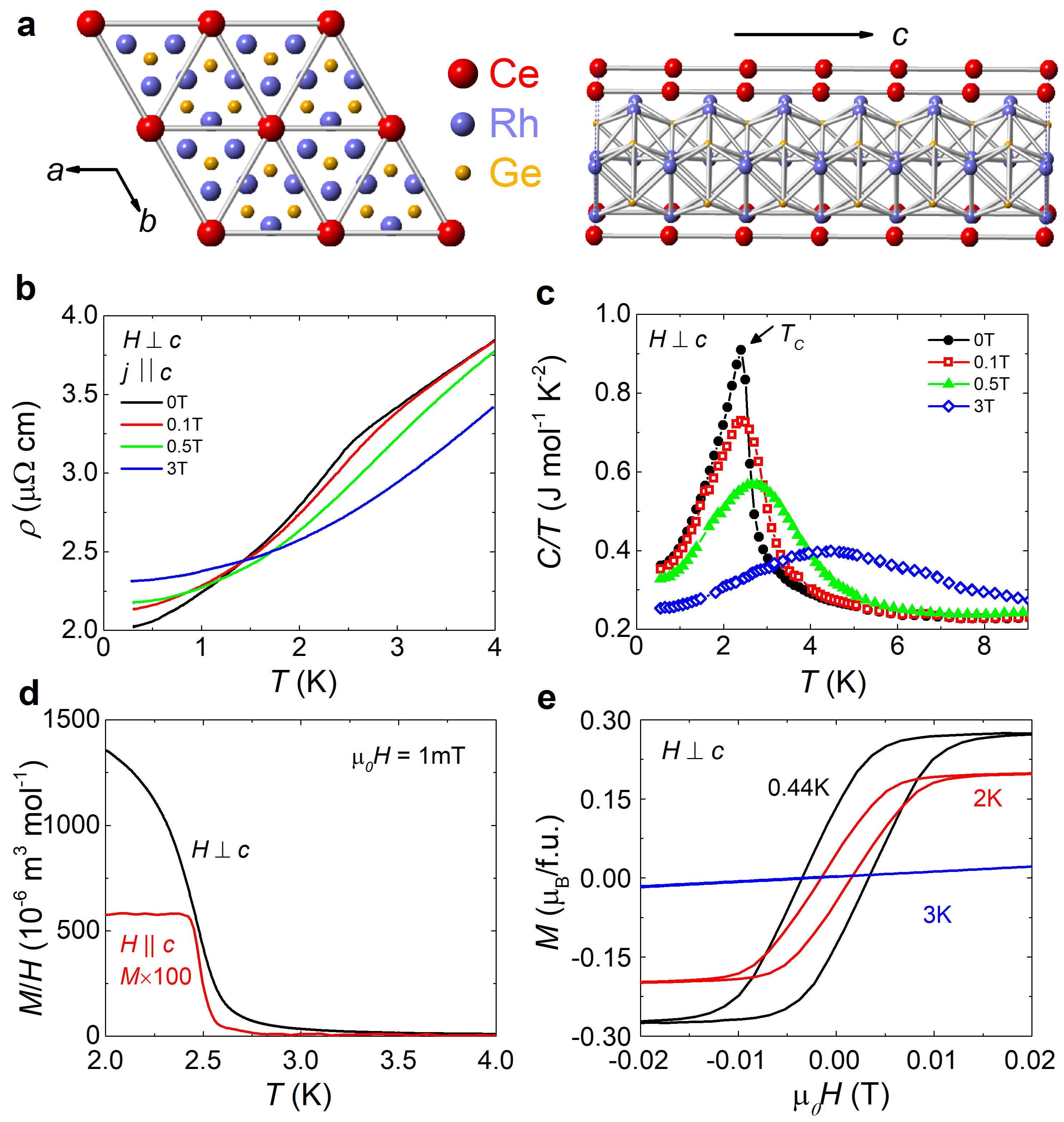}
\end{center}
	\caption{ \textbf{Crystal structure and physical properties of CeRh$_6$Ge$_4$ at ambient pressure.} \textbf{a,} Crystal structure of CeRh$_6$Ge$_4$, where the red, blue and yellow atoms denote Ce, Rh, and Ge respectively. Left panel shows the structure perpendicular to the $ab$-plane, where the Ce atoms have a hexagonal arrangement, while the right side displays perpendicular to the chain direction ($c$-axis).   The  \textbf{b,}  resistivity $\rho(T)$, and \textbf{c,}  specific heat as $C/T$ vs~$T$ of CeRh$_6$Ge$_4$ are also displayed, in both zero-field and various fields applied within the $ab$~plane. \textbf{d,}  Temperature dependence of the magnetization of CeRh$_6$Ge$_4$ as $M/H$ in a field of 1~mT applied both along the $c$~axis and in the $ab$~plane. \textbf{e,} Low field magnetization loops for fields within the $ab$-plane at  three temperatures.  Below $T_{\rm C}$, these exhibit hysteresis loops typical of FM order, while at 3~K no hysteresis is observed.}
   \label{Fig1}
\end{figure}

\begin{figure}[h]
\begin{center}
 \includegraphics[width=0.45\columnwidth]{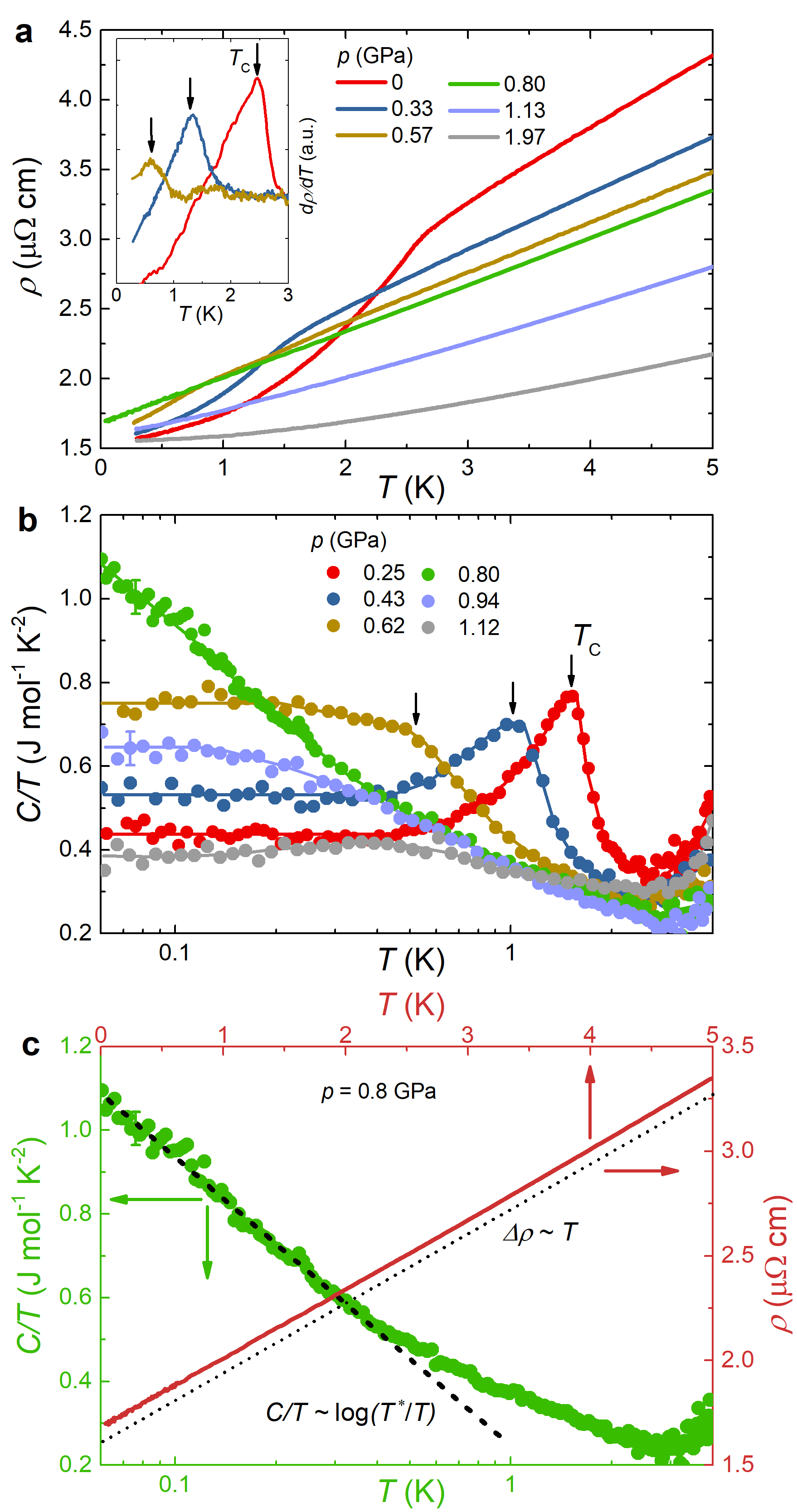}
\end{center}
	\caption{{\textbf{Pressure evolution of ferromagnetism  in CeRh$_6$Ge$_4$ and strange metallic behavior at the quantum critical point.}}  \textbf{a,} Resistivity of CeRh$_6$Ge$_4$ under various hydrostatic pressures. The FM transition is suppressed by pressure, and is no longer observed at $p_c=0.8$~GPa (green line).  The inset shows the derivative of $\rho(T)$ at lower pressures, where the peak position corresponds to $T_{\rm C}$. \textbf{b,}  Specific heat of CeRh$_6$Ge$_4$ under  hydrostatic pressures, where the bulk FM transition is suppressed with pressure, as indicated by the vertical arrows showing the position of $T_{\rm C}$. For clarity, not all the data points are displayed. The error bars shown are representative of the scattering of the data at low temperature. A crossover to FL behavior at low temperatures can  be observed either side of $p_c$,  where $C(T)/T$ flattens. \textbf{c,} $\rho(T)$ and $C(T)/T$ at  $p_c=0.8$~GPa.   $\rho(T)$  exhibits linear behavior extending from 5~K, down to at least 40~mK (dotted line), while $C(T)/T$ continues to increase with decreasing temperature, exhibiting a  $\sim{\rm log}(T^*/T)$ dependence (dashed line). }
   \label{Fig2}
\end{figure}

\begin{figure}[h]
\begin{center}
 \includegraphics[width=0.45\columnwidth]{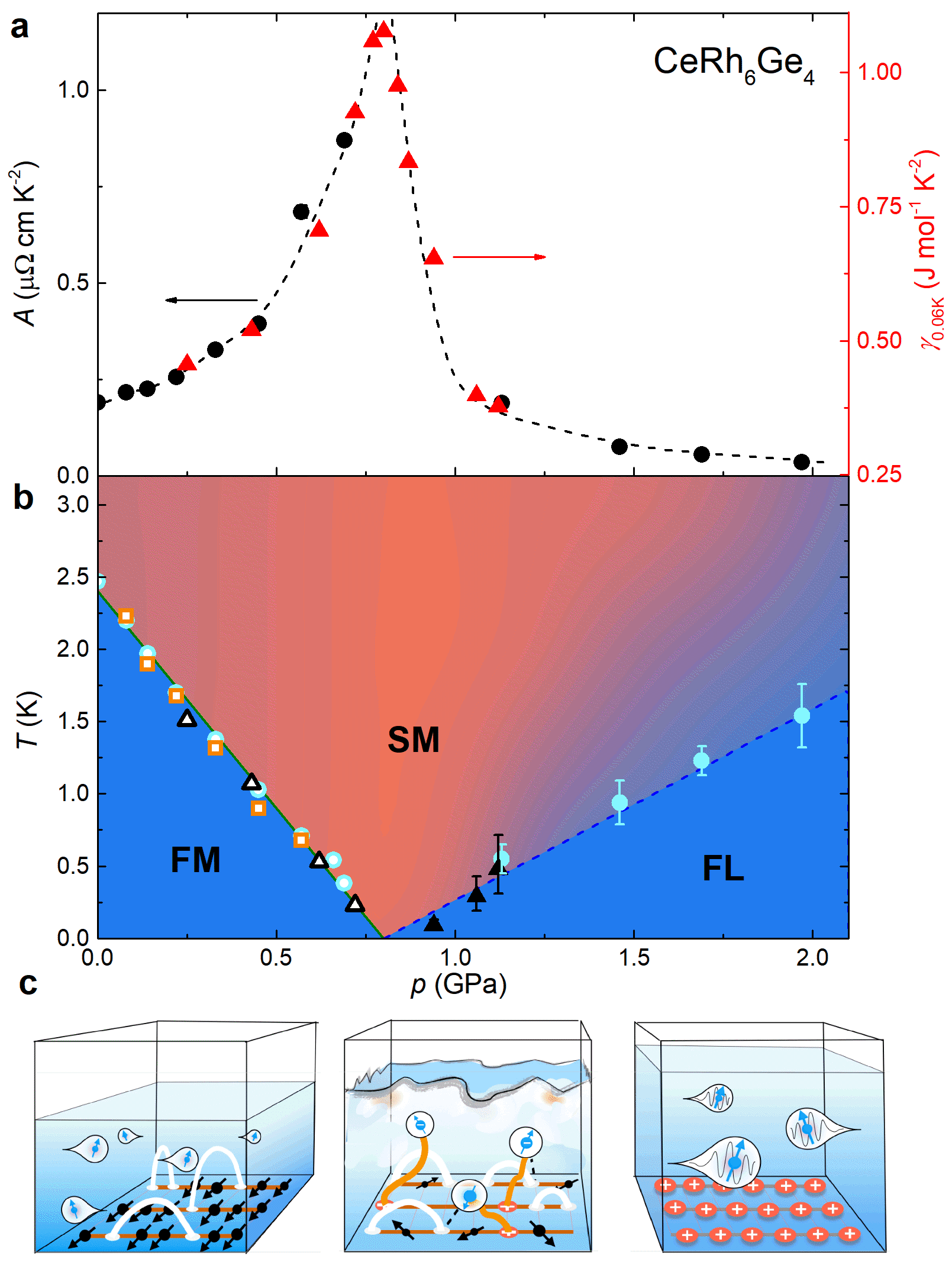}
\end{center}
	\caption{\textbf{Phase diagram of CeRh$_6$Ge$_4$ under pressure.} \textbf{a,}  Pressure dependence of the $A$-coefficient of the $T^2$ term from the resistivity and Sommerfeld coefficient $\gamma$ (as $C/T$ at 60~mK), which show a pronounced maximum near the QCP.  \textbf{b,} Temperature-pressure phase diagram of CeRh$_6$Ge$_4$, where the open circles,  triangles and squares denote $T_{\rm C}$ derived from the resistivity, specific heat (dc method), and ac heat capacity (Fig.~S5), respectively. The corresponding solid symbols mark $T_{\rm FL}$, the temperature below which FL behavior occurs. The FM transition is suppressed by pressure until the system reaches a QCP at $p_{\rm c}\approx0.8$~GPa. Below $T_{\rm C}$, and at higher pressures below $T_{\rm FL}$,  FL ground states develop. These phases (shaded blue) are separated by a fan-shaped region of non-Fermi liquid behavior, where around $p_{\rm c}$ there is a SM phase (shaded red) showing a linear in temperature resistivity and logarithmic  $C/T$ vs~$T$.   \textbf{c,} The schematic representation of different phases. In the ordered phase (left), most of the spins are ordered in the plane, whereas some have RVB bonds. The Fermi surface is small, as 	represented by the volume of the conduction sea. In the PM FL phase (right), all the spins are `ionized' to form heavy-electrons that expand the Fermi sea. A background of positively charged singlets are left behind. At the QCP (center), the system is in dynamical critical equilibrium, where the moments are fluctuating and the Kondo screening by the conduction electron competes with RVBs for the entanglement. In this region, critical fluctuations strongly scatter the conduction electrons.}
   \label{Fig3}
\end{figure}

\end{document}